\documentclass{article}

\usepackage{microtype}
\usepackage{graphicx}
\usepackage{subcaption}
\usepackage{multirow}
\usepackage{booktabs} 

\usepackage{hyperref}

\newcommand{\seq}[1]{\begingroup\ttfamily\footnotesize
\hyphenchar\font=`\- 
\relax #1\endgroup}

\usepackage[preprint]{icml2026}

\usepackage{amsmath}
\usepackage{amssymb}
\usepackage{mathtools}
\usepackage{amsthm}

\usepackage[capitalize,noabbrev]{cleveref}

\theoremstyle{plain}

\theoremstyle{definition}

\theoremstyle{remark}

\usepackage[textsize=tiny]{todonotes}

\icmltitlerunning{Few-Shot Synthetic Accented Speech for ASR Fine-Tuning}

\begin{document}

\twocolumn[
  \icmltitle{Few-Shot Synthetic Accented Speech for ASR Fine-Tuning: \\
    What Helps and When?}



  \icmlsetsymbol{equal}{*}

\begin{icmlauthorlist}
\icmlauthor{Yurii Halychanskyi}{uiuc,ncsa}
\icmlauthor{Nimet Beyza Bozdag}{uiuc}
\icmlauthor{Mark Hasegawa-Johnson}{uiuc}
\icmlauthor{Dilek Hakkani-T\"ur}{uiuc}
\icmlauthor{Volodymyr Kindratenko}{uiuc,ncsa}
\end{icmlauthorlist}

\icmlaffiliation{uiuc}{University of Illinois Urbana-Champaign, USA}
\icmlaffiliation{ncsa}{National Center for Supercomputing Applications, USA}

\icmlcorrespondingauthor{Yurii Halychanskyi}{yuriih2@illinois.edu}

  \icmlkeywords{accented ASR, synthetic speech, data augmentation, phoneme perturbation}

  \vskip 0.3in
]



\printAffiliationsAndNotice{}  

\begin{abstract}
Synthetic accented speech is a promising way to improve automatic speech recognition (ASR) when real accented recordings are scarce. We ask what makes such data useful for ASR fine-tuning: target-accent phoneme edits that expose the recognizer to accent-specific pronunciations, or random phoneme perturbations that act as augmentation in phoneme space. In a few-shot TTS pipeline, we compare LLM-generated accent edits with matched-rate random substitutions and oracle controls using ground-truth accented phonemes and prosody. Random substitutions recover much of the ASR gain: LLM target-accent edits improve over random by only a small margin, ground-truth phonemes stay close to the random baseline and nearly converge with it as the synthetic ASR fine-tuning set grows larger, and adding ground-truth prosody yields only a modest further gain. Mixing synthetic with real accented speech also stabilizes low-resource fine-tuning, but a fixed synthetic budget can later dilute the information in real data, showing that the real--synthetic ratio matters.
\end{abstract}

\section{Introduction}

ASR systems often underperform on accented speech, with accent-related performance gaps documented across benchmarks and bias analyses \cite{Koenecke2020RacialDisparitiesASR,Feng2021QuantifyingBiasASR,AESRC2020}. Existing work addresses this problem from two directions. Model-centric approaches adapt the recognizer itself, for example through accent embeddings, adversarial objectives, or meta-learning \cite{Jain2018AccentEmbeddingsMTL,Sun2018DATAccentedASR,Das2021AdversarialTransferAccents,Winata2020CrossAccentMAML}. Data-centric approaches instead augment training with accented speech generated by voice conversion or TTS \cite{Zhang2022MitigatingBiasNonNativeVC,Klumpp2023SyntheticCrossAccent,Nespoli2024ZeroShotTTSAugmentedASR}. While effective, both directions often require minutes to hours of accented speech \cite{zhou2023accentedtexttospeechsynthesislimited,10317526,Nespoli2024ZeroShotTTSAugmentedASR}, motivating few-shot techniques for settings where only a handful of accented utterances are available. Few-shot accented synthesis is difficult because accent variation spans several factors, including segmental pronunciations, prosody, and speaker-correlated acoustics \cite{AccentBox2024ZeroShotAccentGeneration,prove_accent_with_timbre1,prove_accent_with_timbre2}. Prior work models accents through local/global representations, pronunciation rules, or multilingual resources \cite{Zhou2024MultiScaleAccentModeling,Goronzy2004PronunciationVariantsLexicon,Schaden2003ForeignAccentedRules,MacST2023,Karakasidis}; these approaches improve synthesis, but can still depend on substantial accented data, handcrafted rules, or multilingual corpora.

In this work, we study few-shot synthetic accented speech as a resource for ASR fine-tuning\footnote{Audio samples, the full LLM prompt, and code are available at \url{https://claussss.github.io/few_shot_accent_synthesis_demo}}. We test two hypotheses about why synthetic accented data helps: \textbf{(i)} target-accent phoneme edits improve ASR by exposing the recognizer to accent-specific pronunciation patterns; \textbf{(ii)} random phoneme perturbations improve ASR by acting as noise augmentation in phoneme space and increasing robustness to pronunciation variation. We use an LLM as the practical structured-edit generator: unlike oracle accented phonemes or prosody, it can propose target-accent pronunciation changes for new source utterances from general phonetic knowledge and a few examples, without requiring ground-truth accented phonetic annotations for every sentence. We compare these LLM edits with matched-rate random substitutions, and use ground-truth accented phonemes and prosody as diagnostic oracle bounds. We also study the budget question that arises when real and synthetic accented speech are combined: how much synthetic data helps at a given real-data budget, when it stabilizes fine-tuning, and when it begins to dilute the information carried by real examples. Finally, we examine how little target-accent speech is needed to build the synthetic generator itself.

Our contributions are:
\begin{itemize}
\vspace{-0.4em}
\item Comparing matched-rate random substitutions, LLM target-accent edits, and oracle inputs reveals that phoneme perturbation alone explains much of the ASR fine-tuning gain. LLM target-accent edits and ground-truth accented phonemes both stay close to the random baseline and nearly converge with it at larger budgets, while adding ground-truth prosody yields only a modest further improvement.

\item Mixing synthetic and real accented speech sharply reduces sensitivity to which real utterances are sampled at low real-data budgets, leading to substantially lower run-to-run variance. However, using much more synthetic than real speech can later dilute the benefit of additional real data; the crossover point differs by accent, highlighting the importance of choosing the real--synthetic ratio.

\item Supporting analyses show that ASR trained on one synthetic speaker transfers to other speakers of the same accent, and that the synthetic set can be created from as little as $K{=}3$ target-accent references while LLM editing remains nearly unchanged even with no in-context examples.

\end{itemize}

\section{Method}
\label{sec:method}

Our pipeline starts from a Standard American source utterance, its transcript, and a small set of target-accent reference utterances. We build on the phoneme-conditioned TTS acoustic model of \citet{daft_exprt}. We represent phonemes using the ARPAbet symbol set. For each phoneme $i$, we associate prosodic controls $(d_i,p_i,e_i)$, where $d_i$ is duration in frames using a 256-sample hop at 22.05 kHz, approximately 11.6 ms per frame, $p_i$ is mean log-$F_0$ in log-Hz over voiced frames, and $e_i$ is mean log-energy. These controls are extracted from source speech using transcript-based phoneme alignments rather than predicted internally. We reuse source phoneme-level prosody to preserve natural prosodic diversity from real speech and broaden the synthetic data distribution. To encourage the decoder to follow these externally supplied controls, we add auxiliary pitch and energy reconstruction losses against the extracted frame-level contours during pretraining.

The pipeline introduces target-accent information in two stages. First, we perform target-speaker decoder adaptation: the TTS decoder is fine-tuned on the target-accent speaker's reference utterances, using speaker and style embeddings extracted from the same target speaker. This shifts the acoustic renderer toward the target speaker and accent before any symbolic editing: from a few reference utterances, the decoder can learn speaker-level and accent-correlated realization patterns, so the same phoneme--prosody inputs are rendered closer to how the target speaker produces them. After adaptation, the system generates only the adapted target speaker's voice. Second, an LLM edits the source phoneme sequence to simulate target-accent pronunciations while preserving phoneme--prosody alignment. Insertions, deletions, splits, and merges are allowed; prosody is modified only when needed to keep the sequence aligned. For example, the Standard American realization of \emph{will}, \seq{W IH1 L | d:10,7,7 | p:5.3,5.3,5.2 | e:0.8,3.6,3.1}, may be edited to the Indian-English pronunciation \seq{V IH1 L}; because the sequence length is unchanged, the phoneme-level prosody is copied directly from the source. In-context examples contain paired source and target-accent phoneme--prosody sequences.

\begin{figure}[t]
    \centering
    \includegraphics[width=0.78\columnwidth]{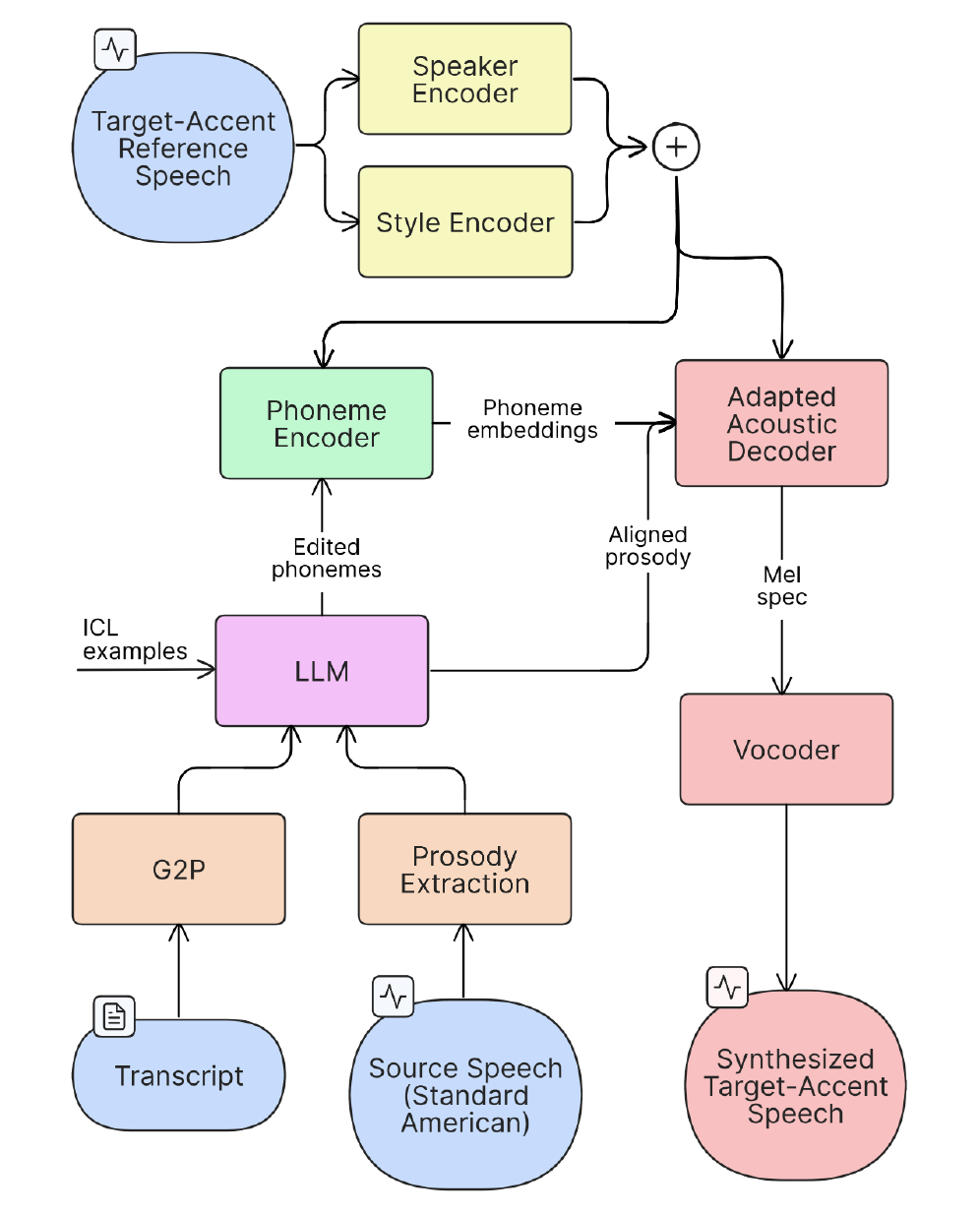}
    \caption{Few-shot synthetic accented speech pipeline used for ASR fine-tuning.}
    \label{fig:diagram_overview}
\end{figure}

\section{Experimental Setup}
\label{sec:setup}

The TTS backbone is pretrained on LJSpeech and the English subset of ESD, both Standard American English only \cite{ljspeech,esd}. Accented speech comes from L2-ARCTIC; CMU Arctic speaker CLB supplies matched Standard American source utterances \cite{l2arctic,cmuarctic}. We evaluate Indian English (TNI, RRBI, SVBI) and Korean English (HKK, YDCK, YKWK). For each accent, one adaptation speaker (TNI or HKK) supplies reference speech for embedding extraction, decoder adaptation, LLM prompting, and synthetic-data generation. Unless otherwise stated, we use $K{=}10$ reference utterances (about 36 seconds), disjoint from evaluation data.

\noindent\textbf{Conditions.}
\textbf{American TTS} uses neither adaptation nor phoneme editing. \textbf{Adapt-only} applies target-speaker decoder adaptation without symbolic edits. \textbf{Adapt+LLM} adds LLM-generated target-accent phoneme edits. \textbf{Adapt+Random} is the perturbation control, replacing uniformly sampled phoneme positions with uniformly sampled ARPAbet phonemes at the same average edit rate as the LLM for each accent: 19\% for Indian English and 35\% for Korean English. \textbf{Adapt+GT-phoneme} replaces source phonemes with accented sequences derived from L2-ARCTIC perceived phoneme labels (PPLs), while retaining aligned source prosody, and \textbf{Adapt+GT-phoneme+prosody} additionally uses prosody extracted from the corresponding real accented audio and aligned to the accented phoneme sequence. \textbf{Real} uses real accented fine-tuning utterances; \textbf{Real+Synth} combines real utterances with LLM-generated synthetic utterances.

\noindent\textbf{Evaluation.}
Acoustic evaluation uses 50 held-out synthesized utterances per speaker (150 total per accent) and reports Whisper-small WER as an intelligibility proxy, UTMOS for naturalness, and SpeechBrain accent-embedding cosine similarity (AccSim) \cite{whisper,utmos,speechbrain}. ASR experiments fine-tune wav2vec~2.0 Base with CTC \cite{Baevski2020Wav2vec2}. ASR scaling varies the number of fine-tuning utterances $N \in [1,500]$. Synthetic-only uses $N$ synthetic utterances, while Real+Synth uses $N$ real utterances plus a fixed set of 500 LLM-generated synthetic utterances. All ASR results are evaluated on real speech from all three speakers per accent (500 utterances each; 1,500 total) and averaged over 7 runs. Oracle comparisons use a separate annotated Indian-English set containing 96 utterances. Additional implementation details for acoustic features, TTS pretraining, adaptation, alignment, LLM prompting and model choice, and ASR fine-tuning are provided in Appendix~\ref{app:impl}.

\noindent\textbf{Supporting analyses.}
For cross-speaker transfer, ASR models fine-tuned on 500 synthetic utterances from the adaptation speaker are evaluated on the other speakers of the same accent. For synthetic-data sample efficiency, we vary $K \in \{1,3,5,10,15\}$ on Indian English (TNI). In each sweep, one component uses only K reference utterances while the remaining components stay fixed at 15: decoder adaptation utterances, speaker-embedding utterances, style-embedding utterances, or LLM in-context examples. The LLM sweep also includes K=0.

\begin{table}[t]
    \caption{Accent synthesis quality under different input conditions. Results are pooled across the three speakers of each accent.}
    \label{tab:accent_synthesis_quality}
    \centering
    \scriptsize
    \setlength{\tabcolsep}{3pt}
    \begin{tabular}{l l r r r}
        \toprule
        \textbf{Accent} &
        \textbf{Condition} &
        \textbf{WER (\%)}$\downarrow$ &
        \textbf{UTMOS}$\uparrow$ &
        \textbf{AccSim}$\uparrow$ \\
        \midrule
        \multirow{6}{*}{Indian English}
          & American TTS              & 6.4  & 3.78 & 0.27 \\
          & Adapt-only                & 11.7 & 2.70 & 0.69 \\
          & Adapt + LLM               & 14.8 & 2.63 & 0.72 \\
          & Adapt + Random            & 47.2 & 2.31 & 0.68 \\
          & Adapt + GT-phon.+pros.    & 20.5 & 2.58 & 0.77 \\
          & Real accent               & 8.6  & 3.89 & 0.86 \\
        \midrule
        \multirow{6}{*}{Korean English}
          & American TTS              & 7.3  & 3.72 & 0.32 \\
          & Adapt-only                & 11.9 & 2.63 & 0.61 \\
          & Adapt + LLM               & 33.8 & 2.51 & 0.61 \\
          & Adapt + Random            & 93.4 & 2.12 & 0.58 \\
          & Adapt + GT-phon.+pros.    & 21.6 & 2.65 & 0.62 \\
          & Real accent               & 14.1 & 3.81 & 0.72 \\
        \bottomrule
    \end{tabular}
\end{table}

\begin{figure}[t]
    \centering
    \includegraphics[width=1.0\columnwidth]{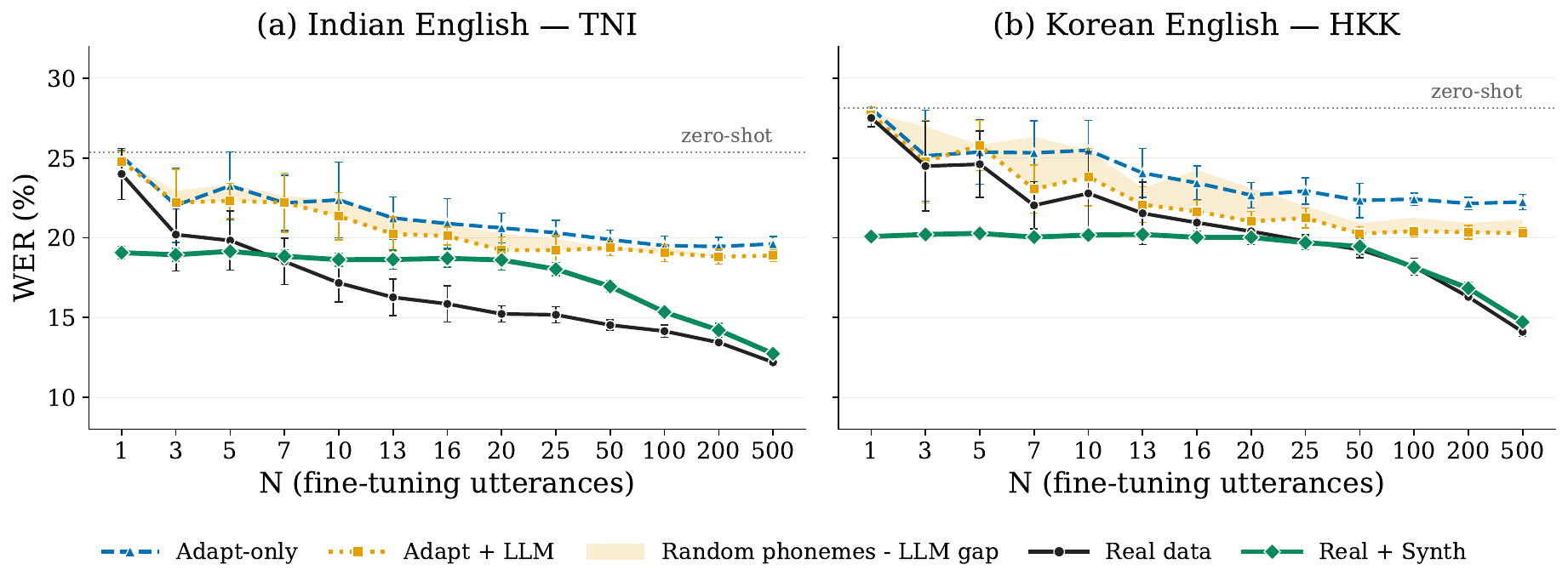}
    \caption{ASR WER versus fine-tuning budget $N$ for Indian (TNI) and Korean (HKK) English. Error bars denote standard deviation over runs. The shaded band shows the gap between Adapt+Random and Adapt+LLM; the band's lower edge corresponds to Adapt+LLM and the upper edge to Adapt+Random. Real+Synth mixes $N$ real utterances with 500 LLM-generated synthetic utterances.}
    \label{fig:asr_scaling}
\end{figure}

\begin{figure}[t]
    \centering
    \includegraphics[
        width=1.0\columnwidth,
        trim=0 0 0 0.34in,
        clip
    ]{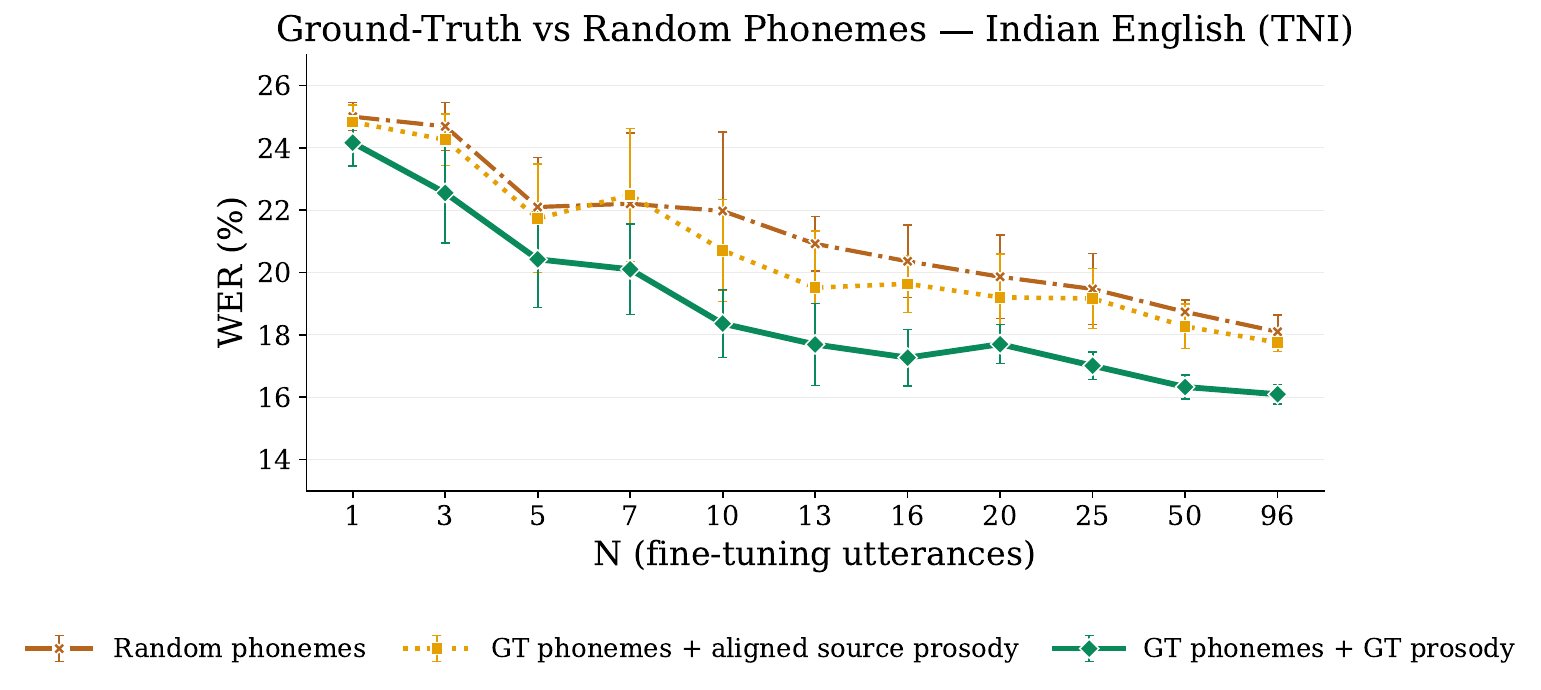}
    \caption{Oracle analysis on an Indian-English subset with ground-truth annotations. GT-phoneme uses accented phonemes with aligned source prosody; GT-phoneme+prosody uses both accented phonemes and forced-aligned target-accent prosody.}
    \label{fig:gt_vs_random}
\end{figure}

\begin{figure}[t]
    \centering
    \includegraphics[width=0.98\columnwidth]{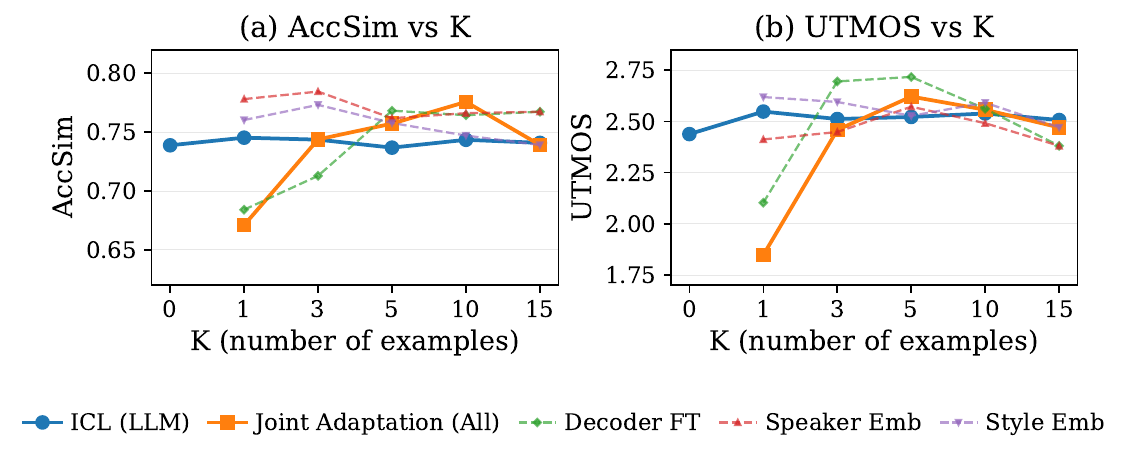}
    \caption{Sample efficiency on Indian English (TNI). (a) Accent similarity and (b) UTMOS versus the number of real target-accent references $K$.}
    \label{fig:fewshot_sweeps}
\end{figure}

\section{Results}

\subsection{What drives ASR fine-tuning gains from synthetic accented speech?}

Table~\ref{tab:accent_synthesis_quality} shows that Adapt-only substantially increases AccSim relative to American TTS for both accents (0.27$\rightarrow$0.69 for Indian; 0.32$\rightarrow$0.61 for Korean), indicating that target-speaker adaptation shifts the rendered speech toward the target-accent acoustics. Adding LLM phoneme edits improves Indian AccSim further (0.69$\rightarrow$0.72) with moderate WER degradation, but does not improve Korean AccSim and raises acoustic WER from 11.9\% to 33.8\%. Korean edits are denser than Indian edits (35\% versus 19\%), suggesting that even after adaptation the decoder struggles to realize the larger symbolic shift intelligibly. Random substitutions are much less acoustically plausible still, reaching 47.2\% WER for Indian English and 93.4\% for Korean English without improving AccSim.

The downstream ASR results tell a different story. Figure~\ref{fig:asr_scaling} shows that Adapt+Random approaches Adapt+LLM for Indian English and is only modestly worse for Korean English. This indicates that a large part of the ASR gain comes from pronunciation-space augmentation: the recognizer benefits from exposure to symbolic variation even when the generated audio is not a faithful target-accent realization. Target-accent edits still help beyond random perturbation, possibly because they remain more linguistically coherent and easier for the adapted decoder to render than arbitrary substitutions, but the margin is small.

Figure~\ref{fig:gt_vs_random} sharpens that interpretation. GT-phoneme remains close to the random baseline across the annotated Indian subset, with a gap that is typically under 1 absolute WER point and nearly vanishes at larger $N$. Even GT-phoneme+prosody reaches only a modest additional advantage over random, about 2 points at the largest budget despite using oracle symbolic and prosodic inputs. Thus, target-accent structure matters, especially at small budgets, but much of the synthetic-only ASR benefit in this setup is explained by phoneme perturbation acting as augmentation.

\subsection{When does synthetic data help alongside real speech?}

Synthetic data provides an additional benefit when combined with real accented speech, but only if the real--synthetic balance is appropriate. In Fig.~\ref{fig:asr_scaling}, Real+Synth improves over real-only fine-tuning at the smallest budgets and sharply stabilizes training: at $N{=}3$, cross-run standard deviation falls from 3.11 to 0.49 for Indian English and from 2.71 to 0.09 for Korean English. The useful range differs by accent. For Indian English, adding 500 synthetic utterances helps mainly at $N{=}3$--7 (19.48\%$\rightarrow$16.81\% at $N{=}3$), after which real-only adaptation quickly overtakes the fixed mixture near $N{\approx}8$. For Korean English, the improvement is larger and persists longer: 20.40\%$\rightarrow$15.83\% at $N{=}5$, with crossover only near $N{\approx}25$.

These crossovers suggest a mixing effect rather than a universal synthetic-data recipe. When real data are extremely scarce, synthetic utterances reduce the dependence on which particular real examples are sampled and lower variance across runs. Once the real set becomes informative, however, a fixed 500-utterance synthetic pool can dilute the stronger signal from newly added real speech. The accent-specific crossover may reflect differences in both base-model mismatch and synthetic-data quality, but the main conclusion is that the synthetic-to-real ratio changes the outcome. We do not optimize mixture weighting here, but the results show that the synthetic-to-real ratio is itself an important design variable.

\subsection{Supporting analyses}

\noindent\textbf{Cross-speaker transfer.}
Although the adapted TTS decoder generates only one speaker, ASR models trained on 500 synthetic utterances from that speaker generalize to unseen speakers of the same accent (Table~\ref{tab:asr_generalization}). This indicates that the synthetic set contains accent-relevant variation useful beyond the adaptation speaker.

\begin{table}[h]
    \caption{Per-speaker ASR generalization results. Values are WER (\%) $\pm$ standard deviation.}
    \label{tab:asr_generalization}
    \centering
    \scriptsize
    \setlength{\tabcolsep}{4pt}
    \begin{tabular}{l r r r r}
        \toprule
        \textbf{Speaker} & \textbf{Zero-shot} & \textbf{Adapt-only} & \textbf{Adapt + LLM} & \textbf{Real} \\
        \midrule
        \multicolumn{5}{c}{\textbf{Indian English}} \\
        \midrule
        TNI  & 25.4 & 19.6$\pm$0.5 & 18.9$\pm$0.4 & 12.2$\pm$0.2 \\
        RRBI & 25.3 & 15.5$\pm$0.3 & 14.6$\pm$0.4 & 10.5$\pm$0.1 \\
        SVBI & 26.1 & 17.4$\pm$0.4 & 17.1$\pm$0.7 & 11.4$\pm$0.3 \\
        \midrule
        \multicolumn{5}{c}{\textbf{Korean English}} \\
        \midrule
        HKK  & 28.1 & 22.2$\pm$0.5 & 20.3$\pm$0.4 & 14.1$\pm$0.3 \\
        YDCK & 20.5 & 15.3$\pm$0.4 & 13.2$\pm$0.2 & 11.6$\pm$0.2 \\
        YKWK & 21.5 & 15.6$\pm$0.6 & 13.9$\pm$0.2 & 12.3$\pm$0.3 \\
        \bottomrule
    \end{tabular}
\end{table}

\noindent\textbf{Synthetic-data sample efficiency.}
Figure~\ref{fig:fewshot_sweeps} studies how much target-accent reference data is needed to create the synthetic set. Under joint adaptation, quality improves sharply from $K{=}1$ to $K{=}3$ (AccSim 0.67$\rightarrow$0.74; UTMOS 1.85$\rightarrow$2.46) and then stays near plateau, so three reference utterances are sufficient in this setup. Decoder fine-tuning is the most sample-sensitive sweep, whereas speaker and style embeddings change only mildly across $K$. The LLM curve is nearly flat from $K{=}0$ to 15, suggesting that pronunciation editing can rely largely on pretrained accent knowledge and reasoning even without in-context examples.

\section{Conclusion}

Few-shot synthetic accented speech improves ASR fine-tuning largely through robustness to pronunciation-space perturbation. In our experiments, much of the gain comes from phoneme-space augmentation, while structured target-accent edits and oracle prosody provide only modest additional benefit beyond random perturbation. In mixed training, synthetic data is valuable because it reduces variance when real data are extremely scarce, yet a synthetic-heavy mixture can dilute increasingly informative real data; choosing the real--synthetic ratio is therefore central. Supporting analyses show that single-speaker synthetic training transfers within accent and that the generator can operate from as little as three reference utterances, with LLM editing remaining robust even without in-context examples. Future work should extend these controls to more accents and learn better strategies for mixing real and synthetic speech.

\section*{Acknowledgements}
This research used the DeltaAI advanced computing and data resource, which is supported by the National Science Foundation (award OAC 2320345) and the State of Illinois. DeltaAI is a joint effort of the University of Illinois Urbana-Champaign and its National Center for Supercomputing Applications.

\section*{Impact Statement}

This work studies why, and under what conditions, synthetic accented speech can improve ASR fine-tuning for accented speakers. The intended impact is to support more inclusive speech recognition systems by reducing performance gaps for speakers whose accents are underrepresented in existing datasets. By analyzing when synthetic data helps, when it mainly acts as pronunciation-space augmentation, and when it should be balanced with real accented speech, the paper aims to make accent-focused ASR adaptation more reliable and data-efficient. At the same time, synthetic accented speech should be used carefully: accent groups are heterogeneous, and simplified synthetic representations may fail to capture the full diversity of real speakers. Responsible deployment therefore requires evaluation across speakers and accents, transparent reporting of data sources and limitations, and attention to possible bias amplification.

\bibliography{example_paper}
\bibliographystyle{icml2026}

\newpage
\appendix
\onecolumn
\section{Supplementary Material}

\subsection{Implementation Details}
\label{app:impl}

\noindent\textbf{Acoustic features and vocoder.}
Mel-spectrograms use 80 mel bins at 22{,}050~Hz, a 1024-point FFT, hop length 256, and mel range 0--8~kHz. Waveforms are synthesized using the universal HiFi-GAN vocoder \cite{hifigan}.

\noindent\textbf{Backbone TTS training.}
The backbone is trained for 72k iterations with Adam (batch size 128). The mel reconstruction loss uses $\ell_1$ and $\ell_2$ terms, each weighted by 1.0. Pitch and energy consistency losses use $\lambda_p{=}1.0$ and $\lambda_e{=}0.2$, and the FiLM regularization weight is $10^{-3}$. Pitch consistency uses a 3-layer 1D convolutional network (kernel size 3, hidden size 256) that predicts frame-level log-$F_0$ from mel-spectrograms. Dynamic speaker statistics are computed from randomly sampled subsets of $M \in \{1,\dots,15\}$ utterances per speaker and refreshed every 2500 training steps.

\noindent\textbf{Few-shot adaptation and alignment.}
We fine-tune the entire acoustic decoder for 600 steps at learning rate 0.001 using mel reconstruction loss only. Text-to-phoneme conversion and phoneme durations use Montreal Forced Aligner English resources with CMU Pronouncing Dictionary lookup and MFA's pretrained English G2P fallback \cite{mfa}. Frame-level pitch is extracted with REAPER; energy is computed from mel-spectrogram magnitude \cite{reaper}. For Adapt+LLM, phoneme edits and required alignment adjustments are generated by an OpenAI GPT-5.1 model via prompt design and in-context examples \cite{openai2024gpt51}. All ASR models are fine-tuned for 15 epochs with AdamW at $3\times10^{-5}$; the wav2vec~2.0 feature extractor is frozen and SpecAugment is disabled.

\end{document}